\documentclass[twocolumn,pre,amsmath,amsfonts]{revtex4}
\usepackage[dvips]{color}
\usepackage[utf8]{inputenc}
\usepackage{pslatex}
\usepackage[T1]{fontenc}
\usepackage{graphicx}
\usepackage{epsfig}
\usepackage{bm,amsmath,amssymb,amsfonts}

\date{\today}
\begin{document}

\title{Asymmetric voltage noise in superconducting tunnel junctions with electromagnetic environment}

\author{Martin \v{Z}onda and Tom\'a\v{s} Novotn\'y}
\affiliation{Department of Condensed Matter Physics, Faculty of Mathematics and Physics,\\
Charles University in Prague, Ke Karlovu 5, 121 16 Praha 2, Czech Republic}

\begin{abstract}
We investigate theoretically the $V$-$I$ characteristics and voltage noise of superconducting tunnel junctions with small critical current 
via the matrix-continued-fraction method. Special attention is paid to the large hysteresis in the $V$-$I$ characteristics 
and to the voltage-noise anomaly observed in preliminary experiments. The current dependence of the voltage noise shows a strong asymmetry between the forward and backward current ramping 
and a discontinuous change of the noise close to its measured maximum occurring at the switching current. We show that both the large hysteresis and the voltage noise anomaly in this current-biased setup are a consequence of the influence of junction's electromagnetic environment. 
Typically, the voltage-noise dependence on the junction current parametrized by the external drive contains a loop which is, however, not observed experimentally because of the implementation of the junction-current ramp. Skipping a part of the loop is responsible for the observed hysteresis as well as the noise anomaly. \newline\newline
\end{abstract}
\maketitle

\section{Introduction}
For small-area Josephson junctions, where the critical current is in the nanoampere range and the capacitance is in femtofarads,
the dynamics of the Josephson phase is strongly influenced by the electromagnetic environment \cite{Joyez}.
This can have profound effects on the measured $V$-$I$ characteristics and voltage noise.  
For example, depending on the realization of the environment the same junction can be in the underdamped regime, 
with a typical hysteretic behavior of the $V$-$I$ curves between the forward and backward source-current ramping, 
but also in the nonhysteretic overdamped regime.

\begin{figure}[h!] 
  \includegraphics[width=0.5\textwidth]{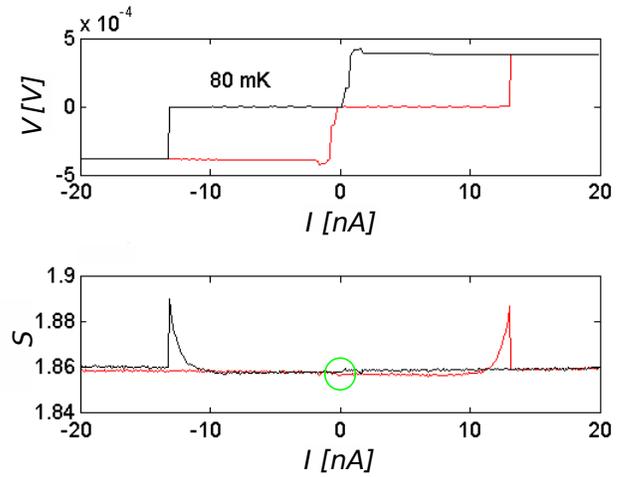} 
 \caption{Experimental dependence of the mean junction-voltage on the bias current (top panel) with the corresponding voltage-noise (in arbitrary units) dependence (bottom panel). Black line represents forward current ramping while the red corresponds to the backward direction. Unpublished data courtesy of P.~Hakonen~\cite{Hakonen}. \label{F1}}
\end{figure}

A recent preliminary experimental study by Pertti Hakonen's group in Helsinki \cite{Hakonen}
focusing on the voltage noise of small-area tunnel junctions revealed a strong 
forward vs.~backward current-ramping mean-voltage and voltage-noise asymmetry and hysteresis (see Fig.~\ref{F1}). 
Furthermore, a discontinuous change of the voltage noise coinciding with the switching current was observed. The purpose of this paper is to show that such surprising effects can be understood as a consequence of junction's electromagnetic environment. 

The organization of our work is as follows. 
First, we introduce in Sec.~\ref{circuit} a simple model circuit close to those used in experiments for obtaining the critical 
current of real junctions and shortly discuss the matrix continued fraction method used to analyze properties of the model circuit. 
In Sec.~\ref{mechanism}, we present the basic ideas of our explanation. Finally, we discuss our numerical results and show that they are in 
qualitative agreement with the experiment in Sec.~\ref{results}.

\section{Circuit and method}\label{circuit}   
\begin{figure}
\centering
\includegraphics[width=0.5\textwidth]{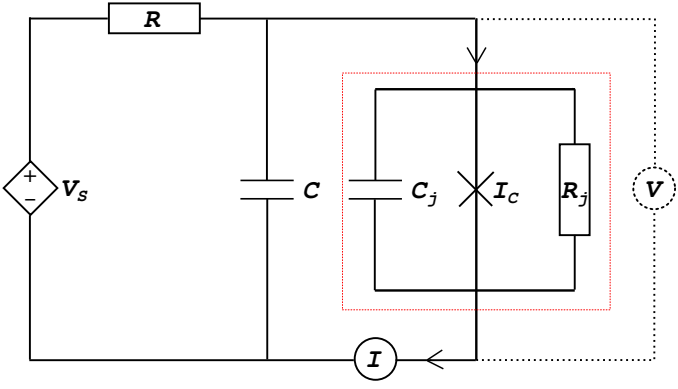}
\caption{Idealized circuit for measurement of the supercurrent 
of Josephson junctions. The red square contains a junction modeled by the RCSJ model. \label{F2}} 
\end{figure}

As it is known the electromagnetic environment can be engineered to place any Josephson junction in the overdamped regime \cite{Steinbach}.
We investigate the influence of such environments on the junction properties in more detail. 
The considered circuit is shown in Fig.~\ref{F2}, where a Josephson junction (red square)
with critical current $I_c$, intrinsic capacitance $C_j$, and 
resistance $R_j$ is biased by a circuit with a capacitor $C$, resistor $R$, and an ideal voltage source $V_s$. 
This circuit is equivalent to a current source $I_s=V_s/R$ in parallel with $R$, $C$, and the junction.
Identical circuit was used to show that the commonly observed reduction of the 
maximum supercurrent in an ultrasmall junction is not an intrinsic junction property, but is due to its electromagnetic
environment \cite{Steinbach}. Its advantage lies in the fact that, in contrast to the usual
current-bias method (applied directly to the junction) giving access only to the positive
differential resistance part of the $V$-$I$ characteristic, all points on the $V$-$I$ characteristic are in principle achievable.
That's because both the current $I$ through the junction (measured by ideal amperemeter $I$) 
and the junction voltage $V$ (measured by ideal voltmeter $V$) are average quantities adjusted to the global drive.

The dynamics of this circuit following from Kirchhoff's laws and Josephson's relations is described by the Langevin equations 
\begin{align}
I(t)&=\frac{V_s-V(t)}{R} - C\dfrac{dV(t)}{dt} + \xi_R(t),\nonumber\\ 
I(t)&=\frac{V(t)}{R_j} + C_j\dfrac{dV(t)}{dt} + I_c \sin{\varphi(t)} + \xi_{R_j}(t),\nonumber\\ 
V(t)&=\frac{\hbar}{2e}\frac{d\varphi(t)}{dt},
\label{E1} 
\end{align}
where two current-noise sources $\xi_R$ and $\xi_{R_j}$  associated with (mutually uncorrelated) thermal fluctuations in resistors $R$ and $R_j$ are assumed to be simple Gaussian white noises satisfying
\begin{equation}
\langle \xi_{R_x}(t)\rangle=0,\hspace{0.2cm} \langle \xi_{R_x}(t_1)\xi_{R_x}(t_2)\rangle=\frac{2k_BT}{R_x}\delta(t_1-t_2).
\end{equation}

It is useful to introduce the equivalent parallel  resistance $R_p=RR_j/(R+R_j)$ and the equivalent parallel capacitance $C_p=C+C_j$. 
Then the quality factor of the circuit reads 
\begin{equation}
Q=\omega_pR_pC_p,
\label{EQ} 
\end{equation}
where $\omega_p=\sqrt{I_c/\varphi_0 C_p}$ is the plasma frequency with $\varphi_0=\hbar/2e$ the reduced flux quantum.
For the purpose of  theoretical analysis it is useful to rewrite Eq.~(\ref{E1}) into the 
dimensionless form by introducing the following dimensionless quantities \cite{Kautz} - junction voltage $ v=\tfrac{QV}{I_cR_p}$, 
junction current $i=\tfrac{I}{I_{c}}$, 
time $\tau=\omega_pt$, temperature $\Theta=\tfrac{k_BT}{\varphi_0I_c}$ (with $k_B$ the Boltzmann constant), driving force $i_s=\tfrac{V_s}{RI_c}$, 
and, finally, the composite Gaussian white noise $\zeta$ with the correlation function   
\begin{align}
\langle \zeta(\tau_1)\zeta(\tau_2)\rangle&=\frac{\omega_p}{I_c^2}\langle (\xi_R(t_1)-\xi_{R_j}(t_1))(\xi_R(t_2)-\xi_{R_j}(t_2))\rangle\nonumber\\
&=2\gamma\Theta\delta(\tau_1-\tau_2),
\end{align}
where $\gamma=1/Q$ is the the damping coefficient.

Using these definitions Eq.~(\ref{E1}) can be reformulated in the dimensionless form ($v(t)=\partial\varphi/\partial \tau$)
\begin{equation}
\frac{\partial v(\tau)}{\partial \tau}= i_s - \gamma v(\tau) - \sin\varphi(\tau) + \zeta(\tau).
 \label{E2}
\end{equation}
This equation is formally identical to the RCSJ model \cite{Stewart}, which is commonly used for description of Josephson junctions 
without the environment and, thus, the methods developed for the RCSJ model can be applied. We have used the matrix continued fraction method \cite[Sec.~11.5]{Risken} for the solution
of the Fokker-Planck equation associated to the Langevin Eq.~(\ref{E2}) for the probability distribution function $W(\varphi,v,\tau)$
\begin{align}
\dfrac{\partial}{\partial \tau}W&=-v\dfrac{\partial}{\partial \varphi}W + 
\dfrac{\partial}{\partial v}\left(\gamma v + \sin\varphi - i_s + \gamma\Theta\dfrac{\partial}{\partial v}\right)W\nonumber\\
&\equiv -L_{FP}W.
\end{align} 
The probability distribution function is obtained numerically
by expanding the $v$-part of $W(\varphi,v,\tau)$ into quantum oscillator basis functions, thus obtaining a tridiagonal coupled system of differential equation, and the $2\pi$-periodic 
$\varphi$-part into the Fourier series \cite[Sec.~11.5]{Risken}. We restrict our analysis here to the stationary case where
the average value of junction voltage $v$ and junction current $i$ can be computed as
\begin{align}
\langle v\rangle &=\int\limits_0^{2\pi}\mathrm{d}\varphi\int\limits_{-\infty}^{\infty}\!\mathrm{d}v\, v W(\varphi,v,\infty),\\
\langle i\rangle &=  i_s-\frac{\gamma}{1+\rho}\langle v\rangle, \label{Ei}
\end{align}
where the environmental parameter $\rho=R/R_j$, and the voltage autocorrelation function reads \cite[Sec.~7.2]{Risken}
\begin{align}
\langle v(\tau)v(0)\rangle =\int\limits_0^{2\pi}\mathrm{d}\varphi\int\limits_{-\infty}^{\infty}\!\mathrm{d}v\, v\,e^{-|\tau|L_{FP}}v W(\varphi,v,\infty).
\end{align}  
Finally, the steady state voltage noise is obtained by time integral of the (connected) voltage autocorrelation function
\begin{align}
S=\int\limits_{-\infty}^{\infty}\mathrm{d}\tau\big(\langle v(\tau)v(0)\rangle - \langle v(\tau)\rangle\langle v(0)\rangle\big).
\label{VN}
\end{align}  

\section{Basic mechanism}\label{mechanism}
\begin{figure}[ht!]
\centering
\includegraphics[width=0.5\textwidth]{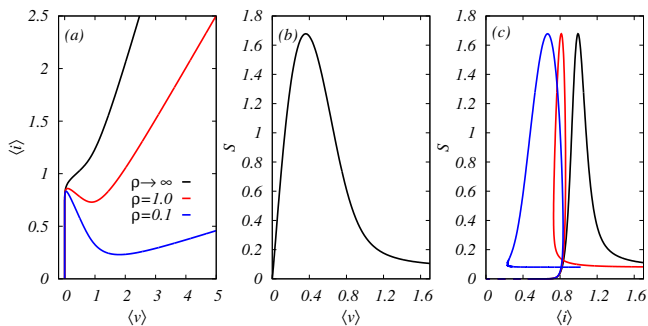}
\caption{\label{F3} Dependence of the junction current $\langle i\rangle$ (a) and the voltage noise $S$ (b) (all curves overlap) on the junction voltage
$\langle v \rangle$ for $\Theta=0.04$, $Q=1$ and different $\rho$'s. (c) Corresponding dependence of the voltage noise $S$ on the junction current $\langle i \rangle$ contains for finite $\rho$ a loop.} 
\end{figure}

In Fig.~\ref{F3}(a) we plot a typical curve of the mean junction current $\langle i\rangle$ 
dependence on the mean junction voltage $\langle v\rangle$ for temperature $\Theta=0.04$,
quality factor $Q=1$,  and different environment parameters $\rho$. Here, limit 
$\rho\rightarrow\infty$ corresponds to the case without the environment. 
For that case the mean current $\langle i\rangle$ is a monotonic function of the voltage 
$\langle v\rangle$. This changes with the introduction of the environment when a local maximum  
and local minimum (decreasing with decreasing $\rho$) of $\langle i\rangle$ develop. Fig.~\ref{F3}(b) shows the voltage dependence of the voltage noise --- 
it turns out that for a given $Q$ this quantity does not depend on the environment details and, thus, all curves for various $\rho$ coincide.   

\begin{figure}[ht!]
\centering
\includegraphics[width=0.5\textwidth]{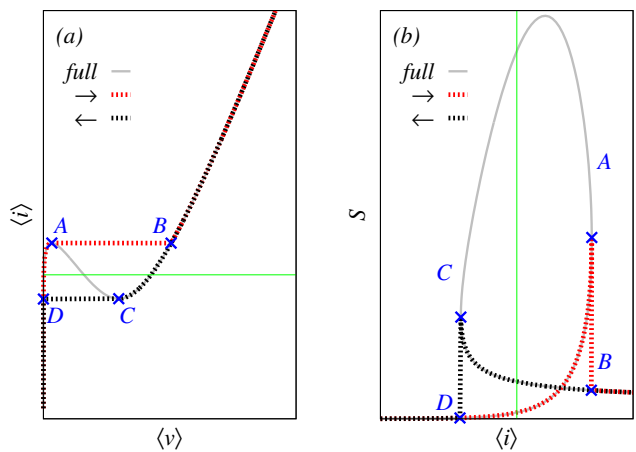}
\caption{\label{F4} (a) Sketch of a typical  current $\langle i\rangle$ vs.~voltage $\langle v\rangle$ characteristic for the RCSJ model of the circuit in Fig.~\ref{F2} 
with shown hysteresis between forward and backward current ramping directions. 
(b) Corresponding picture of a typical voltage noise dependence on the average junction current $\langle i\rangle$ with the hysteresis and discontinuities.} 
\end{figure}  

The non-monotonic dependence of current on the mean voltage has serious consequences for current-biased junctions as illustrated in Fig.~\ref{F4}(a). The mean voltage is not a function of the junction current since in the range between the current minimum and maximum 3 different voltage values can be attributed to the same current (green line in Fig.~\ref{F4}(a)). In practice, when the junction current is ramped up, it will follow the stable branch until it hits the point A when a step back would be required to keep on the average characteristic (grey curve). Instead, the mean voltage will switch to its other stable value at point B (red dots). For the reverse ramp, the analogous situation happens between points C and D (black dotted curve). Depending on the distance between the current extrema the hysteresis can be large.     

The fact that $\langle v\rangle$ is not a function of $\langle i\rangle$ 
has also a profound effect on the voltage noise dependence on $\langle i\rangle$. 
Typically, this dependence (understood as a parametric curve of both quantities parametrized by the mean voltage) contains a loop as shown in Fig.~\ref{F3}(c) and further illustrated in 
detail in Fig.~\ref{F4}(b) by the solid grey line. There the points $A$, $B$, $C$, $D$ correspond to the points in Fig.~\ref{F4}(a). 
For the same reasons causing the hysteresis in $v$-$i$ characteristics one will not observe the whole loop in the noise either 
under the junction-current ramp conditions. The measured noise will leave the loop in the moment 
when a step back in the current bias (negative differential resistance) is needed to remain on the loop.
This coincides with the points $A$ and $C$, respectively, i.e.~with the switching currents for opposite ramp directions.   
Therefore, we suggest this mechanism as responsible for the discontinuous changes of the noise observed in the experiment close to the switching currents (see bottom panel of Fig.~\ref{F1}).
Furthermore, the noise loop is not symmetric and, therefore, the height and the position of the peaks depends on the direction of the measurement as illustrated again by the black and red dotted lines (same color-code as in Fig.~\ref{F1}). As we will show later, for some parameters the lower peak becomes negligible
and in this case we get qualitatively the same noise behavior as in the experiment. To sum up, we have strong evidence that both the large hysteresis and the asymmetric voltage noise are a consequence of the influence of electromagnetic environment on the current-biased junction.

\section{Results and discussion} \label{results}

In this section we show that also other properties of the $V$-$I$ curves and noise are qualitatively in agreement with the experiment. The junction used in the experiment has 
$R_j=7.7k\Omega$, $I_c=19.8nA$, and $C_j=5fF$ and the environment was modeled by $R=0.1k\Omega$ and $C=0.1pF$ in all presented examples.
 
\begin{figure}
\centering
\includegraphics[width=0.5\textwidth]{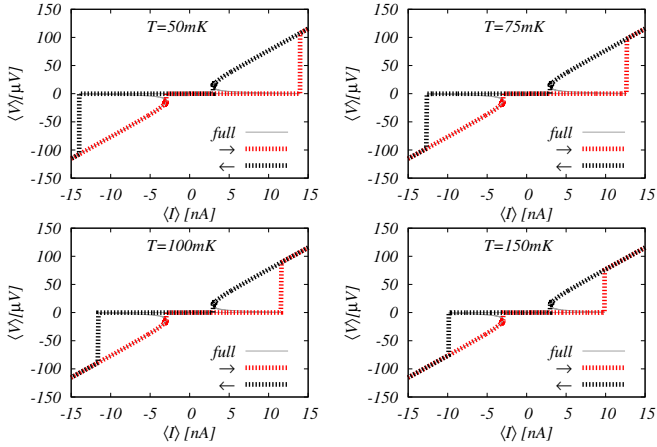}
\caption{\label{F5} Junction voltage $V$ vs.~junction current $I$ for varying temperatures (for other parameter values see the main text).} 
\end{figure}

\begin{figure}
\centering
\includegraphics[width=0.5\textwidth]{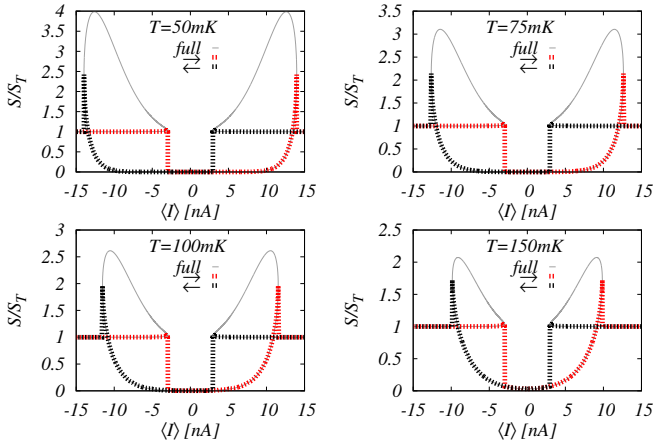}
\caption{\label{F6}  Analogue of Fig.~\ref{F5} for the voltage noise $S$ normalized by $S_{T}=2k_{B}TR_{p}$.} 
\end{figure}

In Fig.~\ref{F5} we plot the $V$-$I$ characteristics for different temperatures
and the current dependence of the voltage noise for the same parameters is plotted in  Fig.~\ref{F6}.
The voltage noise is expressed in units of the thermal Johnson-Nyquist noise $S_T=2k_BTR_p$ of the equivalent parallel resistor $R_p$. 
In all figures the solid grey line represents the full static solution of the circuit Fig.~\ref{F2} and dotted red and black lines show the supposed measured curves when the junction current ramps are considered. From the $V$-$I$ characteristics one can see that the measured hysteresis can by rather large and increases with 
the decreasing temperature. For the chosen parameters the upper drop in the mean junction voltage is approximately $0.1mV$. It is of the same order as in the experiment (Fig.~\ref{F1}).
There is one difference, however. While the asymptotic $V$-$I$ lines are almost horizontal in the experiment (Fig.~\ref{F1}) they follow the $V=R_jI$ line in our theory. 
We believe that this discrepancy would be fixed by  using a nonlinear model for the junction. The nonlinear model gives a better description of real junctions 
for the voltages close to the gap voltage $V_G$ \cite[Sec.~2.3.2]{Likharev}.
A simple piecewise-linear approximation ($R_j(V)=R_{(a)}$ at $|V|<V_G$, $R_j(V)=R_{(b)}$ at $|V|>V_G$ with $R_{(a)}\gg R_{(b)}$) 
should be enough to show that these measured horizontal top and bottom $V$-$I$ lines have their
origin in the almost infinite slope change of $I$ close to the $V=V_G$ known from typical $V$-$I$ characteristics of a Josephson junction \cite[Sec.~2.3.2]{Likharev} 
(note that a vertical line in the $I$-$V$ curve is horizontal in the $V$-$I$ characteristic).

Our calculated voltage noise properties are also in good agreement with the experiment \cite{Hakonen} (experimental data are unpublished yet).    
With decreasing temperature the location of the higher asymmetric noise peak shifts farther away from the zero current and the peak  becomes steeper.
The lower peak is very small even for low temperatures. The very same trends are observed also in the experiment \cite{Hakonen}. 
A closer look at the experimental data in Fig.~\ref{F1} reveals also a small drop in the noise close to $\langle I\rangle=0$ (marked with the green circle).
The realistic environment present in the experiment was apparently more complex than the one assumed here and, thus, quantitative differences are expected, yet our qualitative explanation  should be valid also for more realistic environments.     

\begin{figure}
\centering
\includegraphics[width=0.5\textwidth]{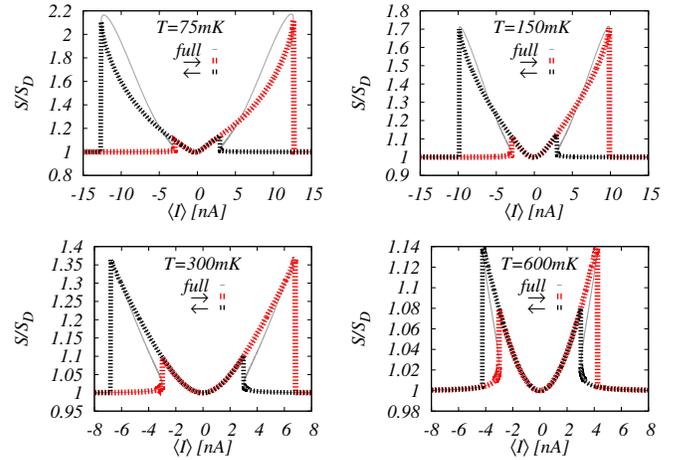}
\caption{\label{F7} Voltage noise $S$ normalized by the equivalent noise $S_D=2k_BTR_D$ of junction's differential resistance vs.~the mean junction current $I$ for wide range of temperatures (for other parameter values see the main text). Normalized-noise unitary value at zero junction current reflects the fluctuation-dissipation theorem.} 
\end{figure}

An important question is the origin of the voltage noise. As demonstrated in Fig.~\ref{F7} we are not dealing exclusively with the thermal noise associated with the combination of the junction differential resistance $R_D=dV/dI$ and resistance $R$. Fig.~\ref{F7} shows the voltage noise $S$ divided by the $S_D=2k_BTR_{Dp}$, where $R_{Dp}=RR_D/(R+R_D)$. For $\langle I\rangle =0$ the ratio $S/S_D$ approaches one (as expected from dissipation-fluctuation theorem) but it exceeds one in the region where the hysteresis of 
$V$-$I$ curves is observed. The origin of this extra noise is in the double-valued nature of the voltage (dichotomous switching) in this regime as can be deduced from the previous RCSJ model studies \cite{Voss} as well as from the fact that the maximum of the noise is decreasing with increasing temperature (typical for dichotomous noise).
 
In conclusion, there are several possible methods for verification of our hypothesis. 
Depending on the details of the measurement the true nature of the $V$-$I$ characteristic could be revealed by using the voltage ramp instead of 
the junction-current ramp. The other possibility is to use a larger number of measurements and slower current ramping.

\begin{acknowledgments}
We thank Aurelien Fay and Pertti Hakonen for sharing their experimental data with us before publication and for useful discussions. We also thank Gabriel Niebler and Tero Heikkil\"{a} for valuable input at the early stages of this work. We acknowledge support by the Czech Science Foundation via grant No.~204/11/J042 and by the Charles University Research Center "Physics of Condensed Matter and Functional Materials".     
\end{acknowledgments}

\end{document}